%% file: main.tex
\documentclass[apsprl,twocolumn,superscriptaddress]{revtex4-2} 
\usepackage{graphicx,subfigure}
\usepackage{amsmath,amssymb}
\usepackage{mathtools}
\usepackage{multirow}
\usepackage{textcomp}
\usepackage{float}
\usepackage{xcolor}
\usepackage{ulem}
\usepackage{dsfont}
\usepackage{siunitx}
\DeclareSIUnit\gauss{G}
\usepackage{bibunits}
\usepackage{import}
\usepackage{braket}
\usepackage{comment}
\usepackage{bm}
\usepackage{bbm}
\usepackage[german,english]{babel}
\usepackage{hyperref}
\hypersetup{colorlinks=true,linkcolor=blue,citecolor=blue,breaklinks}

\newcommand{\abs}[1]{\ensuremath{\left\vert{#1}\right\vert}}

\newcommand{\micro}[1]{\ensuremath{\mu\mathrm{#1}}}

\renewcommand{\micro}[1]{\ensuremath \mu\mathrm{#1}}
\newcommand{\um}{\ensuremath \mu\mathrm{m}}

\renewcommand{\vec}[1]{\ensuremath{\mathbf{#1}}}

\let\oldsfdefault\sfdefault
\usepackage[scaled=0.92]{helvet}
\renewcommand{\sfdefault}{\oldsfdefault}

\defaultbibliographystyle{apsrev4-2}

\begin{document}
\normalem	

\begin{bibunit}
\title{An unsupervised deep learning algorithm for single-site reconstruction\\ in quantum gas microscopes}

\author{Alexander~Impertro}\email{a.impertro@lmu.de}
\author{Julian~F.~Wienand}
\author{Sophie~Häfele}
\author{Hendrik~von~Raven}
\author{Scott~Hubele}
\author{Till~Klostermann}
\author{Cesar~R.~Cabrera}
\author{Immanuel~Bloch}
\affiliation{Department of Physics, Ludwig-Maximilians-Universit\"at M\"unchen, Schellingstr. 4, D-80799 Munich, Germany}
\affiliation{Munich Center for Quantum Science and Technology (MCQST), Schellingstr. 4, D-80799 Munich, Germany}
\affiliation{Max-Planck-Institut f\"ur Quantenoptik, 
             Hans-Kopfermann-Strasse 1, D-85748 Garching, Germany}

\author{Monika~Aidelsburger}
\email{monika.aidelsburger@physik.uni-muenchen.de}
\affiliation{Department of Physics, Ludwig-Maximilians-Universit\"at M\"unchen, Schellingstr. 4, D-80799 Munich, Germany}
\affiliation{Munich Center for Quantum Science and Technology (MCQST), Schellingstr. 4, D-80799 Munich, Germany}

\begin{abstract}
In quantum gas microscopy experiments, reconstructing the site-resolved lattice occupation with high fidelity is essential for the accurate extraction of physical observables. For short interatomic separations and limited signal-to-noise ratio, this task becomes increasingly challenging. Common methods rapidly decline in performance as the lattice spacing is decreased below half the imaging resolution. Here, we present a novel algorithm based on deep convolutional neural networks to reconstruct the site-resolved lattice occupation with high fidelity. The algorithm can be directly trained in an unsupervised fashion with experimental fluorescence images and allows for a fast reconstruction of large images containing several thousand lattice sites. We benchmark its performance using a quantum gas microscope with cesium atoms that utilizes short-spaced optical lattices with lattice constant $383.5\,$nm and a typical Rayleigh resolution of $850\,$nm. We obtain promising reconstruction fidelities~$\gtrsim 96\%$ across all fillings based on a statistical analysis. We anticipate this algorithm to enable novel experiments with shorter lattice spacing, boost the readout fidelity and speed of lower-resolution imaging systems, and furthermore find application in related experiments such as trapped ions.

\end{abstract}
\date{\today}

\maketitle


\section{Introduction}
Efficient data processing based on machine learning techniques has found numerous applications ranging from pattern recognition to the classification of quantum many-body phases~\cite{carleo_machine_2019,carrasquilla_machine_2020}. The potential of machine learning algorithms for applications in experimental quantum physics lies in its power to extract information from experimental or numerical data by reducing the available information to few essential characteristics. Examples include the classification of topological phases of matter based on a limited number of experimental observables~\cite{rem_identifying_2019,kaming_unsupervised_2021}, investigations of the phase diagram of the Fermi-Hubbard model based on (spin-resolved) density snapshots~\cite{bohrdt_classifying_2019,khatami_visualizing_2020} or the multi-parameter optimization of experimental cooling techniques for trapping and imaging of atoms~\cite{wigley_fast_2016,tranter_multiparameter_2018}. These examples highlight the potential of machine learning techniques for faster and more accurate data analysis in particular in the presence of noise or experimental imperfections. Machine learning has recently been considered theoretically in the context of quantum gas microscopes in order to relax the stringent experimental requirements for high-fidelity imaging~\cite{picard_deep_2019}. In this work, we demonstrate a novel architecture for reconstructing the optical lattice occupation from fluorescence images and benchmark it using experimental data.

Quantum gas microscopy has facilitated unprecedented levels of control and observation for the study of quantum many-body systems based on neutral atoms in optical lattices~\cite{gross_quantum_2021,bakr_quantum_2009,sherson_single-atom-resolved_2010}. Experimentally, the system is typically probed using fluorescence imaging, where the atoms are pinned in deep optical lattices and scatter fluorescence photons which are recorded on a camera through a high-resolution imaging system. In order to extract complex observables such as counting statistics and (multi-point) correlation functions~\cite{endres_single-site-_2013}, the lattice occupation needs to be reconstructed in a site-resolved fashion. The attainable accuracy of an observable thus directly depends on the accuracy of the reconstruction. The latter is fundamentally limited by two quantities: (1) The signal-to-noise ratio (SNR) of the recorded image, and (2) the ratio $\beta$ between the imaging resolution and the lattice spacing. Defining the resolution according to the Rayleigh criterion, most existing quantum gas microscopes work with resolutions that are close to the lattice spacing or up to a factor 1.5 worse~\cite{bakr_quantum_2009,sherson_single-atom-resolved_2010,cheuk_quantum-gas_2015,edge_imaging_2015,haller_single-atom_2015,omran_microscopic_2015,parsons_site-resolved_2015,yamamoto_ytterbium_2016,mitra_quantum_2018,kwon_site-resolved_2022,yang_site-resolved_2021,li_high-powered_2021}.

\begin{figure*}[t]
\includegraphics[]{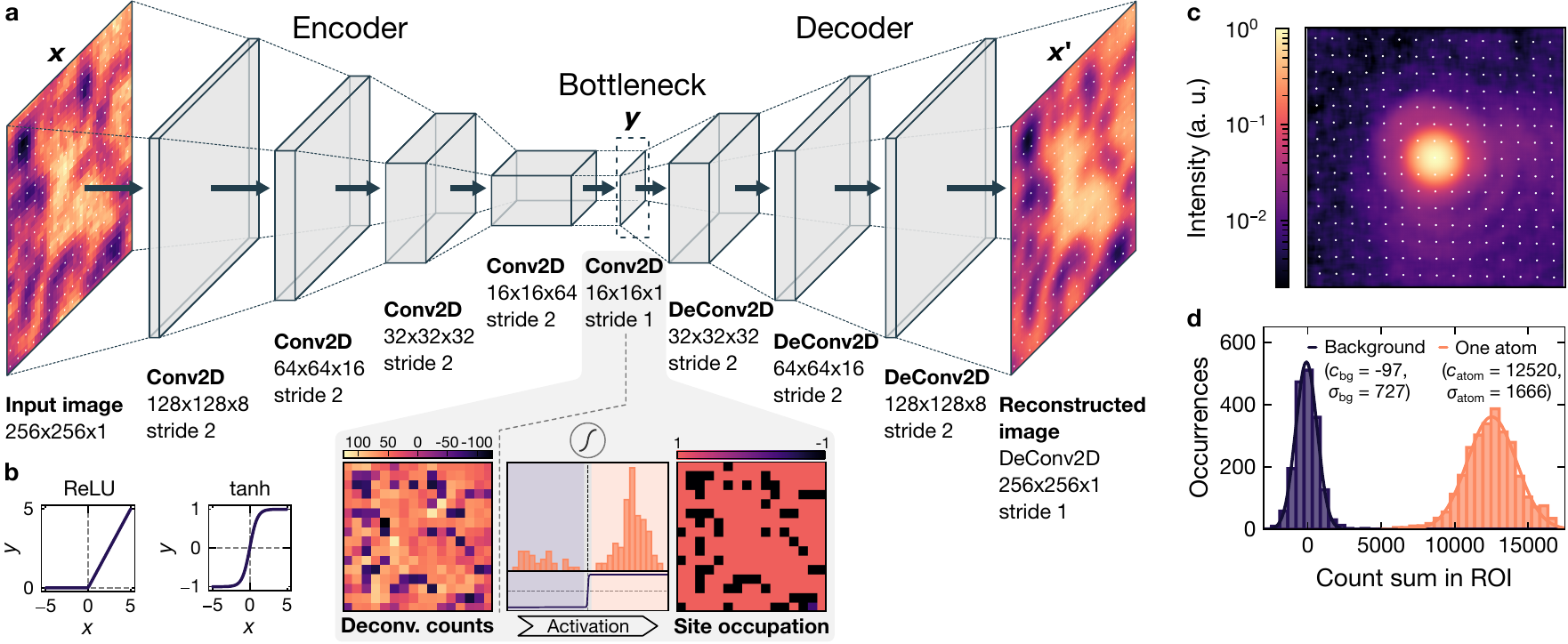}
    
     \caption{\textbf{Network architecture and imaging characterization.} \textbf{a} Regularized convolutional autoencoder architecture, consisting of an encoder and a decoder. The input data passed to the encoder is a lattice section containing $16\times16$ lattice sites ($256\times256$~pixels, white dots mark the lattice sites). This data is transformed by a sequence of five convolution layers, where in each layer a discrete convolution with a set of learned kernels is applied to the respective input. The third dimension in the output shape denotes the number of distinct kernels utilized per layer. Using a step size (\textit{stride}) of two when advancing the kernels through the input during the convolution allows to reduce the image size by a factor of two in each step. At the end of the encoder, the node values of the bottleneck layer after applying a tanh activation function represent the binarized lattice occupancy. The node values before binarization contain the deconvolved counts in each lattice site, which exhibit a bimodal distribution owing to the saturating effect of the tanh function. Subsequently, the occupation matrix is processed by the decoder with the goal of replicating the input image. The decoder network transforms and upsamples the bottleneck layer using four transposed convolution layers. \textbf{b} Rectified Linear Unit (ReLU) activation functions are used in all layers except in the bottleneck and the last decoder layer, which use a tanh activation function. \textbf{c} Measured experimental PSF (averaged over around 2,000 individual PSFs) overlaid with the lattice grid. The central peak spans several sites and we observe long-range asymmetric features extending over the whole region of $16\times 16$ lattice sites. \textbf{d}~Signal-to-noise ratio determined from the count distributions of crops ($\SI{1.6}{\micro\meter}$ crop width) containing exactly zero atoms or one atom. A background image without atoms was subtracted to shift the center of the background peak closer to zero.}
     \label{fig:autoencoder_architecture}
\end{figure*}

In the past, several techniques have been developed to enable high-fidelity reconstruction: One of the first microscope experiments employed an iterative least-squares based approach which is, however, computationally expensive and limited in fidelity for smaller SNR~\cite{sherson_single-atom-resolved_2010, la_rooij_comparative_2022}. Deconvolution with a linear kernel that is constructed to have minimum overlap with atoms on adjacent sites in order to cancel the signal spillover~\cite{greif_site-resolved_2016}, on the other hand, is computationally fast, but limited in performance as it assumes a single kernel that does not depend on the number of neighboring atoms. However, especially for shorter lattice spacings, where the small interatomic separation enhances density-dependent effects during imaging such as superradiance, this assumption is violated~\cite{masson_universality_2022}. Further approaches are based on image restoration techniques such as Wiener filtering~\cite{wiener_extrapolation_1964} or Richardson-Lucy deconvolution~\cite{richardson_bayesian-based_1972,lucy_iterative_1974,omran_microscopic_2015}. Both methods however rapidly decrease in fidelity for $\beta \gtrsim 2$~\cite{la_rooij_comparative_2022}. Moreover, unconstrained deconvolution methods can be improved by adding further information, such as the discrete lattice grid and a sophisticated noise model, as shown in Ref.~\cite{alberti_super-resolution_2016} for a one-dimensional (1D) system.

In general, all these algorithms aim to invert the convolution of the atomic distribution with the point spread function (PSF) of the imaging system, i.e., to realize a deconvolution, which is however ill-conditioned in the presence of experimental noise~\cite{starck_deconvolution_2002}. It has recently been recognized that neural networks can have advantages in solving such inverse problems due to their ability to approximate non-linear relationships~\cite{arridge_solving_2019,genzel_solving_2020}. In addition, their low computational complexity promises a fast evaluation, especially compared to iterative algorithms. With respect to quantum gas microscopy, a reconstruction approach based on supervised neural networks has previously been proposed and benchmarked with simulated data~\cite{picard_deep_2019}. The supervised nature however requires training using simulated fluorescence images. Hence, the performance in reconstructing experimental data will ultimately be limited by the accuracy of the simulation.

In this paper, we present an unsupervised deep learning algorithm for the reconstruction of the lattice occupation from fluorescence images. The unsupervised nature allows us to train the network directly with experimental data, avoiding the need for any simulated training data. We experimentally benchmark the fidelity by analyzing the deconvolved count distributions as well as repeated exposures with data produced by our cesium quantum gas microscope that operates at a rather large resolution-to-spacing ratio of $\beta = 2.2$ (Fig.~\ref{fig:autoencoder_architecture}c). In particular, we find reconstruction fidelities $\gtrsim 96\%$ across all fillings, and we are able to reconstruct large images containing several thousand lattice sites in less than one second. Our scheme enables high-fidelity reconstruction at shorter lattice spacing compared to previous reconstruction algorithms. Besides potentially reducing the technical complexity of the imaging system, this is also important for special lattice configurations such as superlattices, triangular~\cite{yamamoto_single-site-resolved_2020,garwood_site-resolved_2022,mongkolkiattichai_quantum_2022} or state-dependent lattices~\cite{robens_low-entropy_2017}, as well as experiments working with dipolar interactions where close atomic distances are favorable~\cite{baier_realization_2018, chomaz_dipolar_2022}.


\section{Network architecture}
\label{sec:network_architecture}

The problem at hand -- reconstructing the binary lattice site occupation from a recorded fluorescence image -- can be understood in the framework of data analysis as a problem of dimensionality reduction. Here, a high-dimensional noisy input image is transformed into the underlying binary lattice occupation of strongly reduced dimensionality. There exist a variety of dimensionality reduction techniques, with both linear approaches such as principal component analysis (PCA) as well as non-linear methods \cite{van2009dimensionality}. In this work, we leverage the power of artificial neural networks, which fall into the latter category. Their inherent non-linearity allows to efficiently approximate the deconvolution~\cite{arridge_solving_2019} as well as to capture density-dependent effects such as superradiance~\cite{sherson_single-atom-resolved_2010,masson_universality_2022}. Among neural networks, the \emph{autoencoder} is a popular architecture for various dimensionality reduction tasks~\cite{masci_stacked_2011, bengio_representation_2013}. Autoencoders consist of a stacked encoder and decoder network, separated by an interface layer that provides an encoded representation of the input data. The combined network is optimized to replicate the input data, enabling unsupervised training. We restrict the interface layer to a significantly lower dimensionality than the input, forcing the network to learn to extract the salient features of the input data in order to allow information flow through this bottleneck (Fig.~\ref{fig:autoencoder_architecture}a). In our context, the encoder learns the deconvolution from fluorescence image to site occupation, while the decoder learns to simulate a fluorescence image corresponding to this occupation.

We design and implement a regularized convolutional autoencoder architecture that is tailored to our reconstruction task. The network topology is depicted in Fig.~\ref{fig:autoencoder_architecture}a. Beginning from the left, the input layer takes a lattice section containing $16\times16$ lattice sites, corresponding to $256\times256$ pixels. The input image is then subsequently down-sampled by a set of four convolution layers (step size two, ReLU activation, Fig.~\ref{fig:autoencoder_architecture}b), before a final convolution operation (step size one, tanh activation, Fig.~\ref{fig:autoencoder_architecture}b) implements the bottleneck layer with an output matrix of $16\times16$ entries. This concludes the encoder part of the network, whose task is to reduce the input image to an array of site occupations. In the decoder part, the bottleneck layer is up-sampled by three transposed convolution layers (step size two, ReLu activation) and a final transposed convolution layer (step size one, tanh activation) to arrive at one output matrix with the same size as the input image.

In each of these layers, the input data is traversed by a set of kernels with entries learned during training, locally performing a scalar multiplication (see SM Section~\ref{sec_sm:encoder_vis} for details on the convolution operations). After this discrete convolution, a non-linear activation function is applied, producing one output for each kernel. The size of the output in the down-/up-sampling layers is always a factor of two lower/higher than the input resulting from the step size of two. An important aspect of our autoencoder implementation is that it consists exclusively of such convolution layers, which is required to retain spatial order throughout the network. Additionally, we split both encoder and decoder into several convolution layers. Compared to a single convolution layer with kernels spanning the whole input, it was shown that deeper networks with smaller kernels perform significantly better in related applications~\cite{nguyen_optimization_2018, bianchini_complexity_2014, simonyan_very_2015,krizhevsky_imagenet_2017}. While kernels in the first layer act directly on the raw image, subsequent layers are able to learn more complex, high-level features by combining the outputs of the previous layer. Additionally, the stride of two continuously changes the receptive field of kernels such that each layer is sensitive to correlations on different length scales~\cite{springenberg_striving_2015}.

The final operation for determining the site occupancy in the bottleneck layer is the application of the activation function. Considering its sigmoidal shape (Fig.~\ref{fig:autoencoder_architecture}b), a truthful reconstruction of the occupation requires that the input values corresponding to empty and occupied sites saturate the activation function on opposite extremes, respectively. Therefore, the deconvolution implemented by the encoder must lead to a bimodal distribution of count values before binarization. The overlap between the two distributions is a signature of the quality of the reconstruction, which in principle further enables a quantitative estimation of the reconstruction fidelity if the functional form of the two modes is known, as we will detail below.

\begin{figure*}[t]
    \includegraphics[]{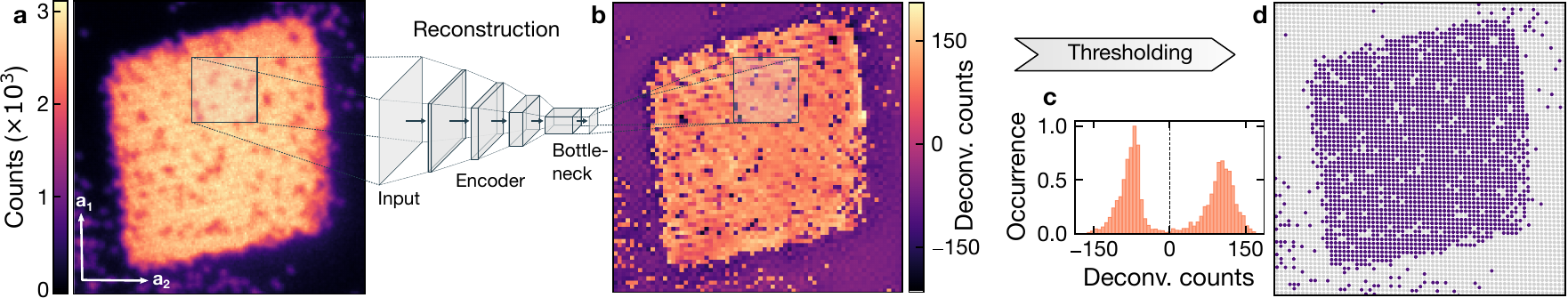}
    \caption{\textbf{Example reconstruction of an experimental image.} \textbf{a} Raw fluorescence image showing a Mott-insulating state ($\sim 2500$~atoms) in a box potential (arrows mark the lattice orientation). Crops containing $16\times16$ sites are extracted from the full image and fed into the encoder part. \textbf{b} The deconvolved site counts are extracted from the bottleneck layer before binarization and reassembled according to their position in the full image. \textbf{c} A histogram of these values reveals a bimodal distribution with a clear separation between empty and occupied lattices sites, allowing to set a threshold for the site occupation. \textbf{d} Applying this threshold finally gives the reconstructed lattice occupation corresponding to the input image. Light gray circles denote empty and dark purple occupied sites.}
    \label{fig:mi_recon_example}
\end{figure*}

The network is trained using a composite loss function
\begin{align}
    \mathcal{L}_\mathrm{tot}\left(\mathbf{x},\mathbf{x'},\mathbf{y}\right)=\mathcal{L}_\mathrm{L1}\left(\mathbf{x},\mathbf{x'}\right)+\frac{\lambda}{N_\mathrm{sites}}\sum_{i=1}^{N_\mathrm{sites}} \left(1-\abs{y_i} \right),
    \label{eq:ae_loss_fn}
\end{align}
where $\mathbf{x}$ and $\mathbf{x'}$ are the input and output images, respectively, $y_i$ are the node values in the bottleneck layer and $N_\mathrm{sites}$ is the number of sites in the image (here, $N_\mathrm{sites}=16^2$). The reconstruction loss ${\mathcal{L}_\mathrm{L1}\left(\mathbf{x},\mathbf{x'}\right) = \sum_\text{pixel}\abs{\mathbf{x}-\mathbf{x'}}}$ is augmented by an additional bottleneck regularization loss, where the regularization strength $\lambda$ determines the relative weight between the two terms. The regularization term ${N_\mathrm{sites}}^{-1}\sum_{i=1}^{N_\mathrm{sites}} \left(1-\abs{y_i} \right)$ penalizes non-binary values in the bottleneck layer, forcing the network to learn a transformation from the input image to the site occupation. Without the bottleneck regularizer ($\lambda=0$), the network would learn to replicate the input image, but the bottleneck would contain a reduced representation of the input image with properties that are difficult to interpret. Our chosen form is one of several possible relations to promote binary node values~\cite{salakhutdinov_semantic_2009}.


\section{Training and application to experimental data}
\label{sec:training_application}

We benchmark the proposed reconstruction algorithm using experimental data. To this end, we prepare a single layer of ultracold cesium atoms in a two-dimensional optical lattice and image the atoms by scattering fluorescence photons on the D2 line with a high-resolution microscope (see SM for further details)~\cite{klostermann_fast_2022, klostermann_construction_2022,von_raven_new_2022}. The minimal distance between atoms during imaging is set by the pinning lattice, which has a spacing of $a = 383.5~\si{nm}$ and the typical resolution according to the Rayleigh criterion is about $850~\si{nm}$. The corresponding resolution-to-spacing ratio is $\beta=2.2$, setting very challenging conditions for the reconstruction. Achieving a high reconstruction fidelity despite the short lattice spacing requires working at a comparably high SNR. Using the definition $\mathrm{SNR} = {(c_{\mathrm{atom}}- c_{\mathrm{bg}})}/{(\sigma_{\mathrm{atom}}+ \sigma_{\mathrm{bg}})}$, where $c_{\mathrm{atom,bg}}$ is the mean and $\sigma_{\mathrm{atom,bg}}$ the standard deviation of the count distribution for atoms and background, respectively, we find an experimental SNR of $5.2$ for an exposure time of $\SI{300}{\milli\second}$~(Fig.~\ref{fig:autoencoder_architecture}d).

In preparation for the reconstruction, we extract the lattice vectors from the Fourier transform of the positions of around 20 dilute atom images. The absolute position of the lattice phase with respect to the image origin is determined for each image by fitting the positions of a few isolated atoms (see SM Section~\ref{supp_sec:lattice_extraction}). Together, this fully defines the lattice grid and ensures that the sites are consistently located at the same positions within the local crops, spanning $16\times 16$ lattice sites. During the training, the network learns that atoms can only be located at this discrete set of positions, which is an important additional information to overcome the resolution limit.

We train the network using an experimental dataset of around 100,000 crops extracted from homogeneous clouds of various average fillings between zero and $n\approx0.98$ (see SM Section~\ref{sup_sec:deterministic_filling} for details on how this data is obtained). The training procedure takes about 100 epochs (full passes through the training set) using the ADAM optimizer (initial learning rate $4\times10^{-4}$)~\cite{kingma_adam_2017}. We performed hyperparameter tuning of the kernel sizes, yielding optimal values of $10\times10$ for the encoder and $22\times22$ for the decoder, respectively. Additionally, the performance of the network is strongly influenced by the choice of the regularization strength~$\lambda$. We optimized this by maximizing the separation in the bimodal distribution of the deconvolved site counts and determined an optimal value of $\lambda_\mathrm{opt} = 0.4$. After successful training, we only use the encoder section of the network for reconstruction.

\begin{figure}[t]
    \includegraphics[]{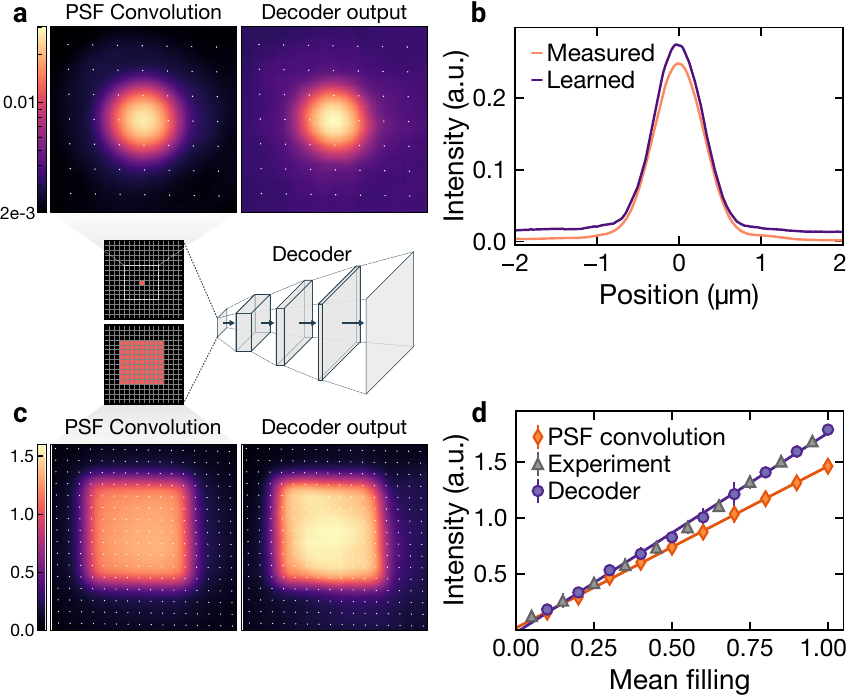}
    \caption{\textbf{Visualization of the decoder.} The transformations learned by the network can be visualized by extracting the decoder part and feeding different occupation matrices into the bottleneck layer. 
    \textbf{a} Analysis of a single occupied site: comparison of the decoder output (right) to the measured PSF (left image, averaged over around 2,000 individual PSFs).  
    \textbf{b} Cut through the center of the 2D images shown in \textbf{a}. 
    \textbf{c}~Analysis of a group of $9\times9$ occupied sites: convolution with the measured PSF (left) and decoder output (right). The latter yields a significantly brighter image. 
    \textbf{d} Mean intensity as a function of filling in regions cropped from experimental data (gray triangles), the decoder output for random occupation matrices at a given filling (purple circles) and a convolution of the same occupation with the measured PSF shown in \textbf{a} (orange diamonds). We only use the central $12\times12$ sites of the crops to avoid edge effects, and the experimental data points are computed from $\sim 1500$ crops per filling bin (same crop size). The lines are linear fits through the respective data points and the error bars denote the standard deviation.}
    \label{fig:ae_decoder}
\end{figure}

The reconstruction process is illustrated in Fig.~\ref{fig:mi_recon_example}, using an example image with a total size of $70\times70$ lattice sites. The raw image (Fig.~\ref{fig:mi_recon_example}a) is first decomposed into crops containing $16\times16$ sites. We advance only by one lattice site between each crop region, such that every site occurs in several analyses at different locations. Each crop is then fed into the encoder part of the network, where we extract the output of the bottleneck layer before binarization via the activation function. The encoder performs the learned deconvolution and transforms the input to a $16\times16$ matrix of deconvolved site counts. These matrices are finally re-assembled according to their position in the original image, averaging the deconvolved counts across overlapping sites (Fig.~\ref{fig:mi_recon_example}b). During the assembly, we only take the central $12\times12$ sites of each crop into account as the border region suffers from a reduced fidelity due to edge effects as a result of missing neighboring sites that are important for the deconvolution. We find that the deconvolution results in a strong enhancement of the contrast of individual holes in the high-density plateau and vice versa of isolated atoms in the outside region. A histogram of all deconvolved site counts (Fig.~\ref{fig:mi_recon_example}c) reveals a bimodal distribution with two well-separated peaks, which we identify as empty and occupied sites, respectively. Since the tanh activation function that is used in the bottleneck layer is symmetric around zero, the correct discrimination threshold is in general zero as well. Applying this threshold to the deconvolved site count matrix finally gives the reconstructed occupation (Fig.~\ref{fig:mi_recon_example}d).

\begin{figure}[t]
    \includegraphics[]{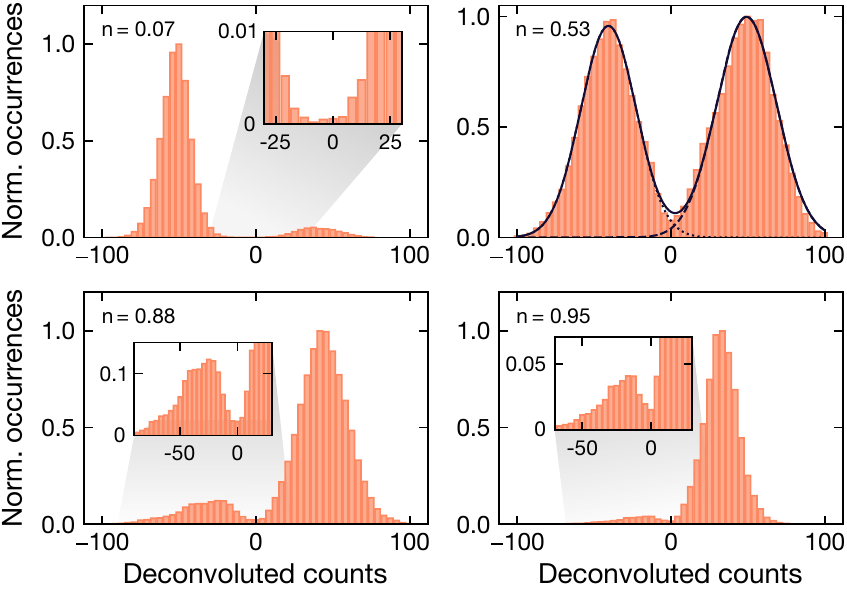}
    \caption{\textbf{Deconvolved site count distributions for different fillings.} The deconvolved site counts are extracted from the bottleneck layer before binarization via the activation function (cf. Fig.~\ref{fig:autoencoder_architecture}). A bimodal distribution distinguishing between empty and occupied sites is found across all fillings. The overlap vanishes for small fillings and increases slightly towards higher fillings. To quantify the fidelity from the overlap, we fit two Gaussians to the data for half filling (solid line), capturing the distributions of empty (dotted line, left peak) and occupied (dashed line, right peak) sites, respectively. Insets show a zoom-in of the overlap region. The filling values were obtained directly from the reconstruction using a threshold of zero, and the occurrences are normalized to the respective maximum in each plot. Each histogram is computed from $\sim 1500$ crops with $12\times12$ sites.}
    \label{fig:amp_histograms}
\end{figure}

A key challenge in employing artificial neural networks for data analysis is to ensure that the network learns a robust, physically reasonable transformation that generalizes to unseen data instead of memorizing the training samples or focusing on coincidental correlations. The symmetric structure of our network enables us to gain more insight into the learned behavior by isolating the decoder part and examining the generated output for varying binary input occupation matrices. We start by setting a single entry of the $16\times16$ occupation matrix to one, expecting that a single PSF appears at the appropriate location in the image generated by the decoder. Figure~\ref{fig:ae_decoder}a and \ref{fig:ae_decoder}b show a comparison between the learned PSF and the measured one, which was extracted from many dilute images (similar to Fig.~\ref{fig:autoencoder_architecture}c). Both the cross-section and the 2D images visualize that the network learns a PSF matching the experimental one in size and shape, with the only difference between them being a small overall offset. As a second test, we investigate to what extent the network is able to approximate density-dependent effects, which are present in our experimental data. To this end, we generate an occupation matrix representing a block of occupied sites (Fig.~\ref{fig:ae_decoder}c). We find that the resulting output image from the decoder is significantly brighter than a simple convolution of the occupation matrix with the measured PSF. To study this quantitatively, we generate random $16\times16$ occupation matrices at various filling fractions and process them using the decoder network. The output is compared to a convolution of the occupation matrix with the measured PSF as well as crops from experimental data ($12\times12$ sites crop size) at the same mean filling. Plotting the mean counts as a function of the filling (Fig.~\ref{fig:ae_decoder}d) reveals a significantly increased brightness for higher fillings in the decoder generated images, which is compatible with the experimental data. This implies that the network is indeed able to capture the density-dependent effects due to superradiance, which in our experiment leads to a $22\%$ higher signal at unity filling.


\section{Performance evaluation}
\label{sec:performance_eval}

\begin{figure}[t]
    \includegraphics[]{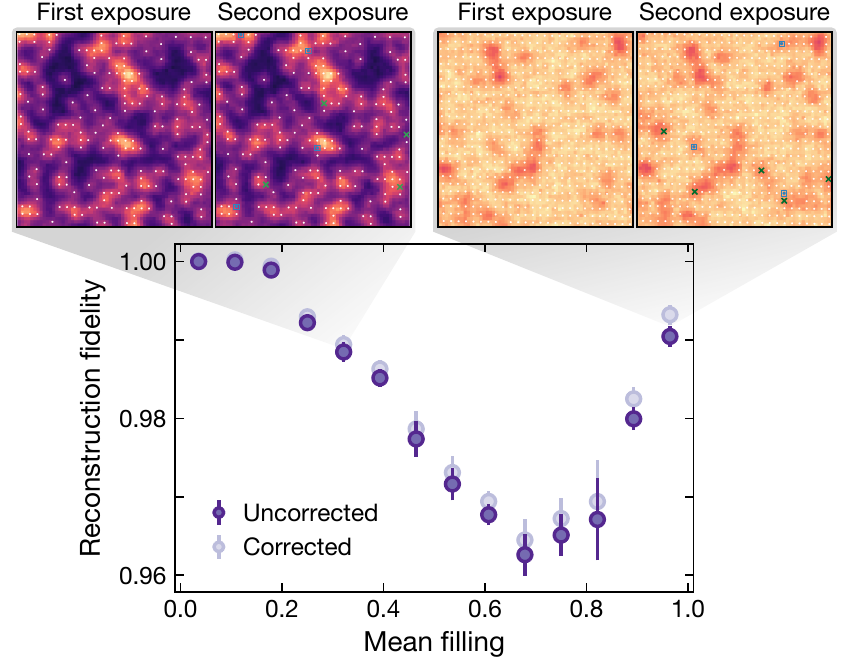}
    \caption{\textbf{Reconstruction fidelity estimation from double-exposure imaging.} By measuring the same many-body state twice, we can estimate the reconstruction fidelity from the difference in the reconstructed occupations via Eq.~(\ref{eq:rho_main}). The uncorrected data is obtained by neglecting hopping and loss, i.e. $p_{\delta}=0$, while the corrected points (shaded) take the independently calibrated value ${p_{\delta}(n)=n\cdot5.9\times10^{-3}}$ into account. Upper panels show two typical images at the indicated mean filling with first (second) exposure on the left (right). The blue squares in the second image indicate sites that changed from unoccupied to occupied, and the green crosses vice versa. Error bars denote the standard error of the mean.}
    \label{fig:multi_img_fidelity}
\end{figure}

In an experimental setting, the reconstruction fidelity is limited by the finite SNR, systematic errors such as spatially inhomogeneous fluorescence as well as hopping and atom loss. A direct measurement of the fidelity in presence of these effects is hindered by the imperfect knowledge of the occupation labels in experimental images. One option would be to benchmark the algorithm on simulated data, but conceiving a simulation that accurately captures all effects present in the experiment is a highly challenging task and beyond the scope of this work. Instead, we present two approaches that allow us to estimate the experimental fidelity directly from the reconstruction process applied to experimental images. 

First, we analyze the distribution of deconvolved site counts by extracting the node values of the bottleneck layer before binarization. Fig.~\ref{fig:amp_histograms} shows the extracted site count distributions for four selected average fillings, where the value for the filling is computed from the reconstruction. We find well-discriminated bimodal distributions across all fillings, with a vanishing overlap at low filling (upper left panel), and a slightly increased overlap towards higher fillings. Without any processing, the distributions of empty and occupied sites would completely overlap due to imaging noise and the spillover of neighboring sites (see SM Section~\ref{sec_sm:unproc_counts}). The effect of the deconvolution is now to separate these distributions again, which is possible up to a certain limit given by the aforementioned experimental imperfections. The remaining overlap can then be used as an approximation to estimate the reconstruction fidelity. In principle, it is expected that the fidelity is maximal for fillings close to zero and unity, and drops to a minimum around half filling due to a maximum possible number of nearest and next-nearest neighbor configurations. In addition, density-dependent effects such as superradiance and the increased effective atom loss due to hopping can reduce the contrast of holes in experiments at high filling. 

A drawback of the non-linear transformation applied by the encoder is that the resulting functional form of the bimodal distribution is a priori not known. This makes it challenging to accurately quantify the reconstruction fidelity. Nevertheless, applying the network to simulated data suggests that the deconvolved distributions are well approximated by normal distributions and do not exhibit strong asymmetries (see SM Section~\ref{supp_sec:simulations} for details). We can therefore estimate the fidelity of the reconstruction process from the overlap region between the two peaks by fitting two Gaussians to the deconvolved counts at half filling (upper right panel). The overlap area corresponds to wrongly classified sites when discriminating via a threshold value, and the reconstruction fidelity is given by the fraction of this overlap area to the total area under the bimodal distribution. From the fit, we determine an experimental reconstruction fidelity for half filling of $\mathcal{F}\sim99\%$.

Second, we quantify the reconstruction fidelity by taking two consecutive images of the same many-body state and comparing the reconstruction results. This method is sensitive to statistical fluctuations of the recorded counts at finite SNR, which limits the overall detection fidelity. We can derive a relation for the reconstruction fidelity $\mathcal{F}$: 
\begin{equation}
\mathcal{F} = \frac{1}{2} \left( 1 + \sqrt{\frac{1-2 \delta}{1-2\,p_{\delta}(n)}} \right),
\label{eq:rho_main}
\end{equation}
where $\delta$ is the probability for finding different reconstruction results on a lattice site between the two images and $p_{\delta}$ accounts for thermal hopping and atom loss, which was calibrated independently (see SM Section~\ref{supp_sec:double_imaging} for a detailed derivation).

Figure~\ref{fig:multi_img_fidelity} shows the measured reconstruction fidelity $\mathcal{F}$ as a function of the average filling $n$. We show data points both with (${p_{\delta}(n)=n\cdot5.9\times10^{-3}}$) and without ($p_{\delta}=0$) the hopping and loss correction. We find a reconstruction fidelity above $99\,\%$ close to unity filling, for $n\lesssim0.2$. A minimum occurs around $n=0.7$, yielding an uncorrected reconstruction fidelity of $96.3(3)\,\%$.

While the double-imaging analysis enables to quantify the reconstruction fidelity, it is primarily sensitive to statistical errors stemming from a finite SNR. However, an analysis of the algorithm with simulated fluorescence images for our experimental parameters showed that we do not expect large systematic errors (see SM Section~\ref{supp_sec:simulations} for details).


\section{Conclusion}

We have presented an unsupervised deep learning algorithm for the reconstruction of the lattice occupation from fluorescence images obtained with quantum gas microscopes. Using experimental images from our cesium experiment, we demonstrated high-fidelity reconstruction of the site occupation using the proposed algorithm in a challenging regime, where the lattice spacing is more than two times smaller than the imaging resolution.
Based on a convolutional neural network that is able to efficiently approximate a wide variety of non-linear relationships, our approach has an inherent advantage in solving the deconvolution problem and capturing density-dependent effects~\cite{sherson_single-atom-resolved_2010,masson_universality_2022}. The autoencoder architecture allows training directly with experimental images, using a dataset with about 200 fluorescence images that can be taken within a few hours in a typical microscope setup. This also obviates the need for a detailed calibration of imaging parameters such as PSF or image noise. The well-optimized tensor operations allow the reconstruction of a full image in less than a second on standard computer hardware. Finally, we have developed and tested novel methods for benchmarking the reconstruction performance, providing a path to estimate the fidelity with experimental data where the true occupation is a priori not known.

Our work enables high-fidelity reconstruction in experiments with short-spaced optical lattices, paving the way for new experiments with exotic lattice configurations~\cite{yamamoto_single-site-resolved_2020,garwood_site-resolved_2022,mongkolkiattichai_quantum_2022}. We anticipate our scheme to remain applicable for even shorter lattice spacings, as long as the SNR is increased appropriately. Beyond the field of quantum gas microscopy, this algorithm can also be applied to other experimental platforms, such as Rydberg-atom arrays~\cite{saffman_quantum_2010,morgado_quantum_2021} or ion trap experiments~\cite{bruzewicz_trapped-ion_2019}.


\section*{Acknowledgments}

The authors would like to thank Christian Schweizer for contributions to the early stages of this experiment and Annabelle Bohrdt for helpful discussions on machine learning architectures. A.I. and T.K. were supported by the Bavarian excellence network ENB via the International Ph.D. Programme of Excellence Exploring Quantum Matter (ExQM). J.F.W. acknowledges support from the German Academic Scholarship Foundation and the Marianne-Plehn-Program. H.v.R. acknowledges support from the Hector Fellow Academy. C.R.C. has received funding from the European Union’s Horizon 2020 research and innovation programme under the Marie Skłodowska-Curie grant agreement No 897142 and the MCQST seed funding (EXC-2111/Projekt-ID: 390814868). We acknowledge funding from the Deutsche Forschungsgemeinschaft (DFG, German Research Foundation) via Research Unit FOR 2414 under project number 277974659 and under Germany’s Excellence Strategy – EXC-2111 – 390814868 and from the German Federal Ministry of Education and Research via the funding program quantum technologies – from basic research to market (contract number 13N15895 FermiQP).

\putbib[references]
\end{bibunit}
\cleardoublepage

\include{supplements}

\end{document}

%% file: supplements.tex
\begin{bibunit}
\onecolumngrid
\begin{center}
 {\large\textbf{Supplementary Information}}
\end{center}
\renewcommand{\theequation}{S\arabic{equation}}
\renewcommand{\thefigure}{S\arabic{figure}}
\newcommand{\h}{\hat{H}}
\newcommand{\U}{\hat{U}}
\newcommand{\Q}{\hat{Q}}
\newcommand{\n}{\hat{n}}
\newcommand{\vj}{\vec{j}}
\newcommand{\va}{\bm{\alpha}}
\newcommand{\tl}{\tilde{L}}
\newcommand{\tth}{\tilde{\theta}}
\newcommand{\dv}{\Delta\tilde{\theta}}
\setcounter{equation}{0}
\setcounter{figure}{0}
\setcounter{section}{0}
\setcounter{secnumdepth}{4}

\vspace{20px}

In this supplementary material we present further details about the experimental setup and describe how we generate fluorescence images of homogeneously distributed atoms at different fillings so that they can be used as training data for the neural network (Section \ref{supp_sec:experimental_details}). We also elaborate on important details of the reconstruction process (Section \ref{supp_sec:reconstruction}), including the extraction of the lattice sites in experimental images and other data pre-processing steps, and demonstrate how the autoencoder-based reconstruction performs with simulated fluorescence images (Section \ref{supp_sec:simulations}). Finally, we present the mathematical details behind estimating the experimental fidelity from taking two images of the same cloud (Section \ref{supp_sec:double_imaging}).

\vspace{20px}

\twocolumngrid

\section{Experimental Details}
\label{supp_sec:experimental_details}

A detailed discussion of the experimental setup used to generate a Bose-Einstein condensate (BEC) in the science chamber has been introduced in Refs~\cite{Klostermann2022, klostermann_construction_2022, von_raven_new_2022}. In the following, we focus on the ingredients needed for the preparation of a Mott insulator (MI) and the fluorescence imaging technique.

\subsection{Setup Overview}
\label{sec:supp_setup}

\begin{figure}[tbh!]
\includegraphics[width=0.99\columnwidth]{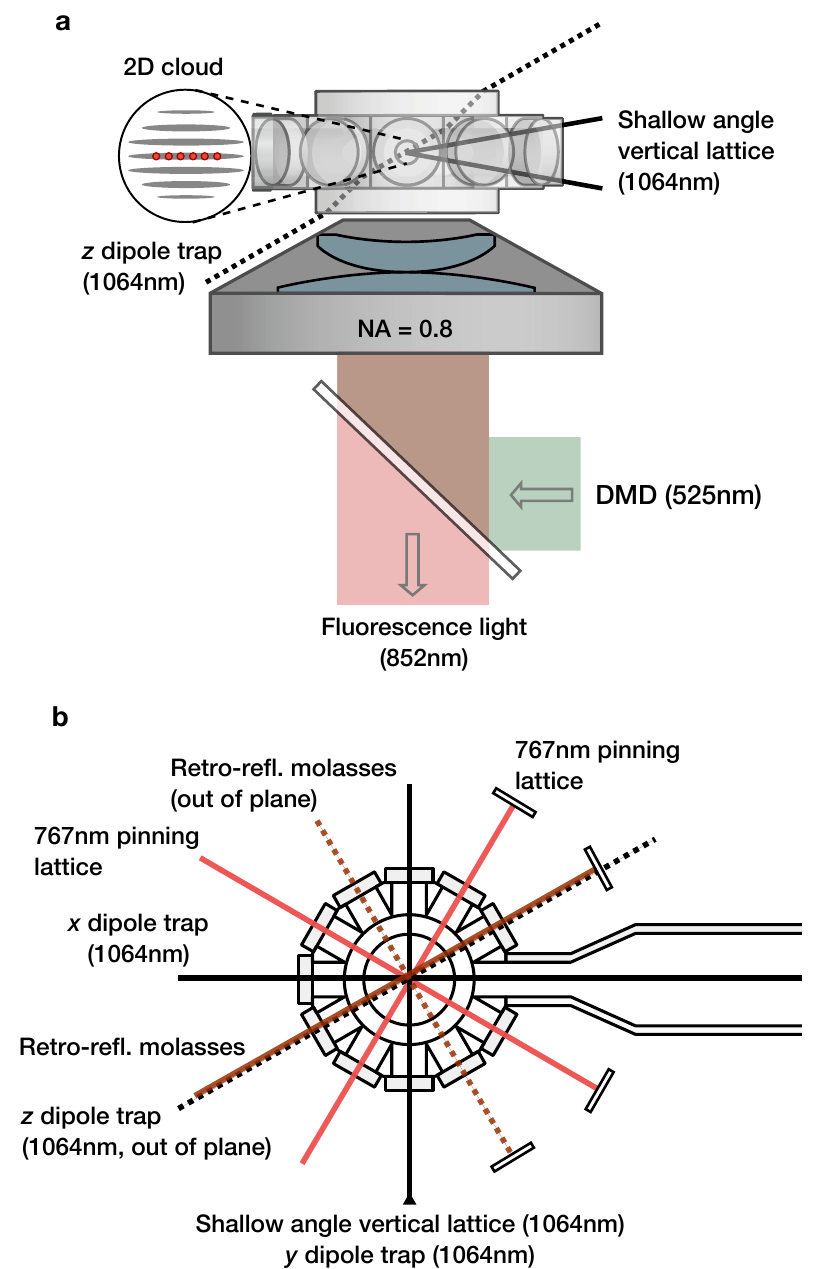}
    
     \caption{\textbf{Experimental Setup.} Dipole traps, lattices and optical molasses in the science chamber, \textbf{a} side view, \textbf{b} top view. Solid lines indicate beam paths in the horizontal plane, dashed lines indicate that the beam path is out-of-plane.}
     \label{fig:supp_setup}
 \end{figure}
 
\emph{Glass cell.--} The science chamber is a glass cell of dodecagon shape with 11 side windows (\SI{12}{\milli\meter}~diameter) as well as a top and a bottom window (\SI{30}{\milli\meter}~diameter). Our high-resolution objective ($\mathrm{NA}=0.8$) is mounted directly underneath the glass cell, optically accessing the atoms through the bottom window (see Fig.~\ref{fig:supp_setup}a).

\begin{figure*}[t!]
\includegraphics[width=1\textwidth]{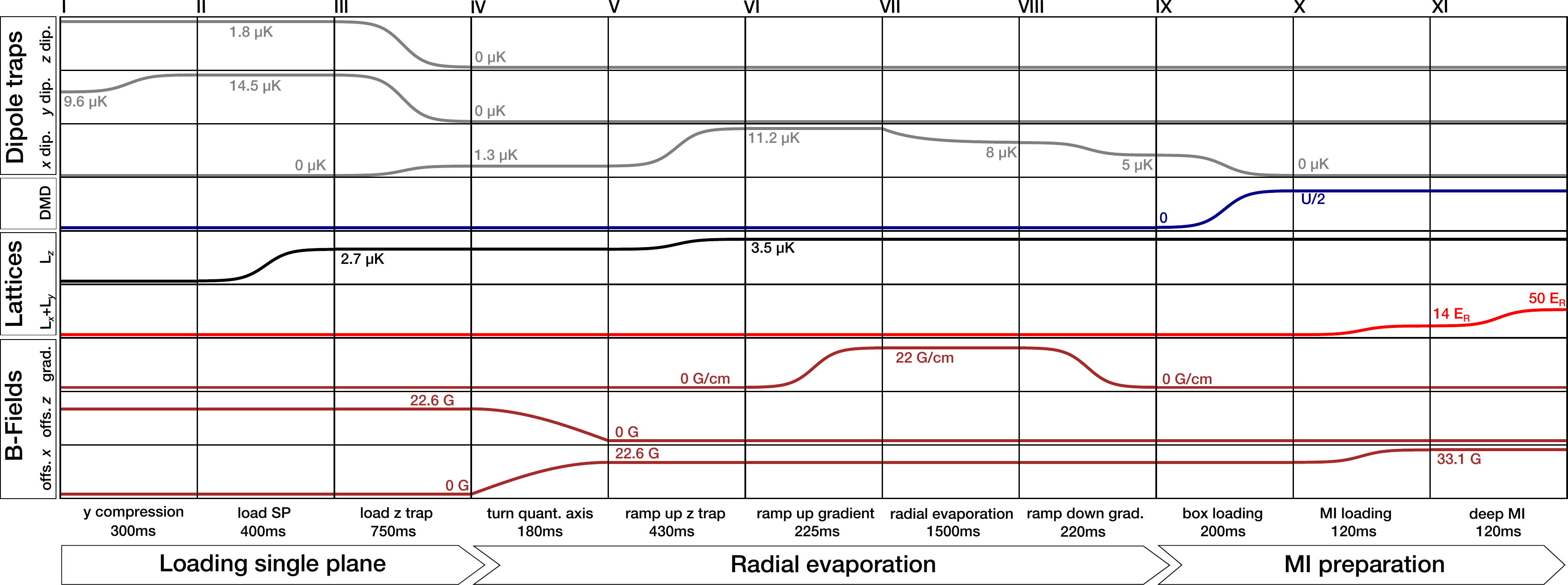}
    
     \caption{\textbf{Mott insulator preparation.} Full experimental sequence for loading the BEC into a single plane of the vertical lattice and turning it into a large Mott insulator with high filling. Note, that the horizontal axis is not to scale.}
     \label{fig:supp_sequence}
 \end{figure*}

\emph{Dipole Traps.--} The science chamber is equipped with three dipole traps and three lattices, providing loading and pinning confinement in all spatial directions. As indicated in Fig.~\ref{fig:supp_setup}b, the $x$ and $y$ dipole traps ($\SI{100}{\micro\meter} \times \SI{100}{\micro\meter}$ and $\SI{370}{\micro\meter} \times \SI{110}{\micro\meter}$ waist, respectively) are perpendicular to each other in the horizontal plane, forming a crossed dipole trap at the position of the atoms at the center of the glass cell. The $z$ dipole trap enters the glass cell under a vertical angle of \SI{60}{\degree} through the top window, closely passing by the objective (Fig.~\ref{fig:supp_setup}a). Its elliptical waist ($\SI{110}{\micro\meter} \times \SI{53}{\micro\meter}$) is designed to create a (projected) round confinement in the horizontal plane. 

\emph{Lattices.--} The atoms are trapped in a single plane of a vertical lattice with a spacing of \SI{8}{\um}. It is the result of two \SI{1064}{\nano\meter} beams ($\SI{170}{\micro\meter} \times \SI{35}{\micro\meter}$ waist) interfering under a small angle of \SI{3.6}{\degree}. The horizontal lattice confinement is provided by two perpendicular retro-reflected beams ($\SI{150}{\micro\meter} \times \SI{45}{\micro\meter}$ waist) with a wavelength of $\lambda=\SI{767}{\nano\meter}$. All three lattices are used both for physics experiments (e.g. to prepare the Mott insulating initial state) and for pinning during fluorescence imaging. The maximum lattice depth is around \SI{400}{\micro\kelvin} in the $x$, $y$ and $z$-direction.

\emph{Optical molasses.--} For fluorescence imaging we perform molasses cooling on the D2 line ($\lambda=\SI{852}{\nano\meter}$) using two retro-reflected molasses beams ($\SI{2}{\milli\meter}$ diameter) and a repumper. The first beam lies in the horizontal plane and shares an axis with the projected beam path of the (out-of-plane) $z$ dipole trap. The second beam enters the glass cell under a vertical angle of \SI{60}{\degree} through the top window (similar to the $z$ dipole trap beam but from a direction that is perpendicular to the first molasses beam). The retro-mirrors of the molasses beams are mounted on piezoelectric actuators which are modulated at around \SI{100}{\hertz} in order to wash out the polarization gradient lattice and achieve a more homogeneous fluorescence signal.

\emph{Imaging Path.--} After passing the objective ($\SI{25}{\milli\meter}$ effective focal length), the fluorescence photons are focused by a tube lens ($\SI{1000}{\milli\meter}$ focal length, Thorlabs ACT508-1000-B) onto a \mbox{sCMOS} camera (Teledyne Photometrics Kinetix). The magnification of this imaging system is 40. For projecting custom potentials onto the atoms in the horizontal plane, we use a DMD (digital micromirror device, Texas Instruments DLP6500, interface by bbs Bild- und Lichtsysteme GmbH) that reflects light generated by a \SI{5}{\watt} multimode laser (Wavespectrum WSLX-525-005-400M-H) running at $\SI{525}{\nano\meter}$. The DMD imaging path demagnifies the mask displayed on the DMD chip by a factor of 160 and projects it onto an area of about $\SI{50}{\micro\meter} \times\SI{90}{\micro\meter}$ in the atom plane. A dichroic mirror is used to separate the light from the DMD going into the objective and the fluorescence light coming from the objective (cf. Fig.~\ref{fig:supp_setup}a).

\emph{Coils.--} The science chamber is surrounded by multiple pairs of coils in close vicinity (not shown in Fig.~\ref{fig:supp_setup}) that allow the generation of offset and gradient fields in all spatial directions. Further, the table on which the experiment is mounted is placed inside a coil cage that is used for compensating environmental magnetic field fluctuations at the position of the atoms (cf. Section~\ref{sup_sec:deterministic_filling}).

\subsection{Mott Insulator Preparation}
\label{sec:supp_mott}

In Fig.~\ref{fig:supp_sequence} we show the full experimental sequence used for preparing a Mott insulator in a box potential with about 2500 atoms in the $n=1$-plateau and up to \SI{98}{\percent} filling. Starting point is the BEC in the crossed dipole trap, formed by the $x$ and $y$ dipole trap beams \cite{klostermann_fast_2022}.

\emph{Single plane loading.--} As a first step, the BEC is loaded into a single plane of the vertical lattice, making the system two-dimensional. We use the $y$ dipole trap to compress the degenerate gas vertically (I) and, subsequently, ramp up the vertical lattice potential (II). The compression allows us to achieve a higher single-plane loading efficiency. We then transfer from the crossed dipole trap into the $z$ dipole trap (III) by ramping down the former and, simultaneously, removing the latter. The $z$ dipole trap provides a more isotropic harmonic confinement and makes it easier to achieve efficient evaporation.

\emph{Radial evaporation.--}  As a next step, we perform in-plane evaporative cooling by applying a gradient in the horizontal plane. We turn the quantization axis in our system (IV) by ramping down the $z$ offset field and, at the same time, ramping up the $x$ offset field, while keeping the total field strength near the three-body-loss-minimum \cite{chinPrecisionFeshbachSpectroscopy2004, weberBoseEinsteinCondensationCesium2003}. In preparation for the evaporation process, we increase the harmonic confinement (V) and apply a horizontal gradient force (VI) which, at this point, is not yet strong enough for pulling the atoms out of the single plane in the horizontal direction. We then evaporate by slowly reducing the horizontal confinement provided by the $z$ dipole trap using an exponential ramp profile (VII). Finally, the gradient is adiabatically turned off and the $z$ dipole trap power is reduced to its final value (VIII). During evaporation any remaining population of neighboring planes of the vertical lattice becomes negligible and we end up with a low-temperature sample in a single plane.

\emph{Mott insulator loading.--} In order to load the ${n=1}$~Mott insulator state into a box potential we start by transferring the evaporation-cooled superfluid from the $z$ dipole trap to a blue-detuned box potential provided by the DMD (IX). The box potential has a $50 \times 50$ sites large quadratic trapping region, delimited by a \SI{4}{\micro\meter} thick wall whose barrier height linearly drops to zero with increasing distance from the trapping region [the slope is $U/(\SI{24}{\micro\meter})$]. Next, we simultaneously ramp up both horizontal lattices in two consecutive ramps, the first one designed to reach the SF-to-MI transition at $U/J=16$ (X) and the second one ending in the deep MI regime with $U/J > 50$ (XI), making sure that the SF-to-MI transition is crossed slowly. Together with the first ramp, we also increase the offset field to \SI{33.5}{\gauss}. Thanks to the broad Feshbach resonance in cesium, this increases the scattering length from $280\,a_0$ to $580\,a_0$ and results in onsite interactions of $U \approx h\times\SI{900}{\hertz}$. The wall height of the box potential is chosen to match $U/2$, such that excess atoms spill out of the box without forming an $n=2$~shell. Those atoms which have spilled over gather in the surroundings of the box due to the remaining horizontal harmonic confinement of the red-detuned vertical lattice beams (see Fig.~\ref{fig:mi_recon_example}a in the main text).

\subsection{Fluorescence Imaging}
\label{sec:suppfluo}

\begin{figure}[tbh!]
\includegraphics[width=0.99\columnwidth]{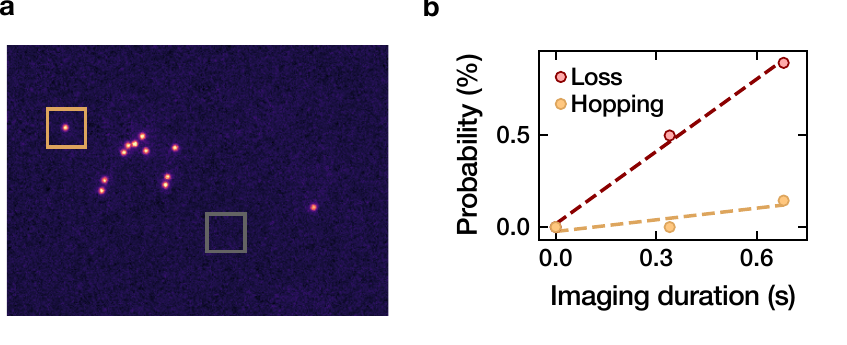}
    
     \caption{\textbf{Imaging characterization.} \textbf{a} Fluorescence image of a dilute cloud of atoms used for analyzing the PSF and determining the SNR by comparing the counts in crops with one atom (orange) and without any atom (gray). \textbf{b} Loss and hopping probability as a function of imaging time, extracted from multiple images of the same dilute cloud of atoms. }
     \label{fig:supp_fluo_imaging}
 \end{figure}

\emph{Sequence.--} After preparing the MI state, we image the atoms by performing molasses cooling on the D2 line ($\lambda=\SI{852}{\nano\meter}$) and collecting the fluorescence photons using an $\mathrm{NA}=0.8$ objective and a sCMOS camera. For imaging, all lattices (both horizontal lattices and the vertical lattice) are quenched to pinning depth. Immediately after pinning, all offset and gradient fields are switched off instantly and the molasses beams are turned on. Their detuning is set to \SI{70}{\mega\hertz} with respect to the $F=4 \leftrightarrow F'=5$ transition. Additionally, a repumper, resonant with the $F=3 \leftrightarrow F'=4$ transition, is switched on to prevent the atoms from occupying the $F=3$ ground state. Shortly (\SI{10}{\milli\second}) after beginning the cooling process, we start exposing the sCMOS camera to the fluorescence light for \SI{300}{\milli\second}.

\emph{Point spread function (PSF) and signal-to-noise ratio (SNR).--}  In order to characterize our imaging resolution, we take several hundred images of very dilute clouds of atoms, as shown in Fig.~\ref{fig:supp_fluo_imaging}a. We analyze these images by cutting crops with exactly zero atoms or one atom inside. The crops containing exactly one atom are used to obtain the average experimental PSF shown in Fig.~\ref{fig:autoencoder_architecture}c. A Gaussian fit yields a width that corresponds to a Rayleigh resolution of about \SI{850}{\nano\meter}. From the count statistics of the crops, as shown in Fig.~\ref{fig:autoencoder_architecture}d, we can determine the experimental SNR (as defined in the main text).

\emph{Hopping and loss rates.--} In order to determine the loss and thermal hopping rates during the imaging process, we image very dilute clouds of atoms (as shown in Fig.~\ref{fig:supp_fluo_imaging}b) three times in a row, with an exposure time of \SI{300}{\milli\second} per image and a small delay of $\SI{40}{\milli\second}$ between the exposures, yielding a total imaging time of \SI{1020}{\milli\second}. In this dilute regime where clustering of atoms is very unlikely, the reconstruction fidelity is assumed to be 1. From comparing the site occupations of the second and third image with that of the first image, we obtain loss and hopping probabilities, as shown in Fig.~\ref{fig:supp_fluo_imaging}b. Using linear fits, we extract a hopping rate of $2.1(1.2)\,\mathrm{mHz}$ and a loss rate of $13(1)\,\mathrm{mHz}$. Note that an atom hopping to a position outside the region of interest (ROI) is also counted as a loss event.

\subsection{Deterministic Preparation of Different Fillings}
 
\label{sup_sec:deterministic_filling}

\begin{figure}[tbh!]
\begin{center}
  \includegraphics[width=0.9\columnwidth]{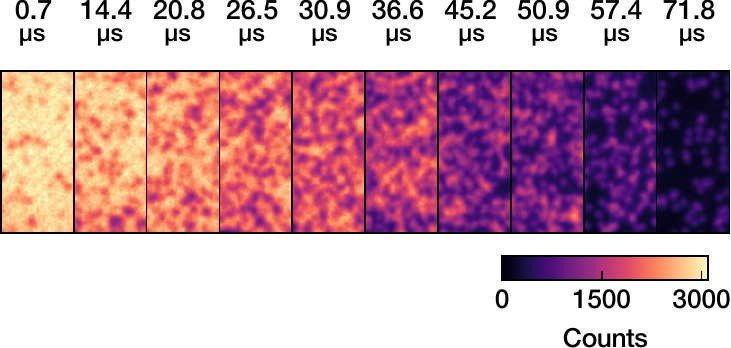}
  \caption{\textbf{Preparation of different fillings.} Cropped fluorescence images of the Mott insulator plateau undergoing Rabi oscillations on the $F=3, m_F=3 \leftrightarrow F=4, m_{F}=4$ transition. We image the $F=3, m_F=3$ state. The microwave pulse duration is varied to cover the first half of the first oscillation period, yielding a variety of different fillings.  }
  \label{fig:supp_boxrabi}
  \end{center}
\end{figure}

For training the neural network and for evaluating its performance as a function of filling, we prepare experimental images with homogeneously distributed atoms at various fillings.  We start by preparing a Mott insulator of about $40 \times 40$ lattice sites. At the end of the MI preparation sequence and before fluorescence imaging, the atoms are in the $F=3, m_{F}=3$ state. In order to reduce the filling from $n=1$ in a controlled fashion, we coherently drive the $F=3, m_F=3 \leftrightarrow F=4, m_{F}=4$ transition at \SI{9.2}{\giga\hertz} using a resonant microwave pulse of variable duration. All atoms transferred to the $F=4, m_{F}=4$ state are subsequently removed by a blowout pulse resonant with the $F=4 \leftrightarrow F'=5$ transition. Fig.~\ref{fig:supp_boxrabi} shows the fluorescence signal from a cropped region inside the MI plateau as a function of the pulse duration. The filling decreases gradually during the first half of the first Rabi oscillation period, as expected. In order to be able to drive the  $F=3, m_F=3 \leftrightarrow F=4, m_{F}=4$ transition coherently, we rely on very stable magnetic field conditions, as the transition frequency shifts with \SI{2.45}{\mega\hertz}$/$\SI{}{\gauss}. We reduce the environmental magnetic field fluctuations from up to $\SI{5}{\milli\gauss}$ to below $\SI{100}{\micro\gauss}$ using a commercially available electromagnetic interference (EMI) compensation system (IDE MK5). The magnetic field noise is measured using sensors placed in the direct vicinity of the glass cell and compensated by applying a feedback current to a coil cage surrounding the experiment \cite{von_raven_new_2022}. 

\section{Reconstruction scheme}
\label{supp_sec:reconstruction}

\subsection{Lattice extraction}
\label{supp_sec:lattice_extraction}
\emph{Extraction of lattice vectors.--} The real space lattice vectors are extracted from images with low filling fraction. For this, isolated atoms are fitted with a two-dimensional Gaussian in order to determine their position with sub-pixel accuracy. A two-dimensional Fourier transform of these coordinates reveals peaks corresponding to the reciprocal lattice vectors, from which the lattice orientation and spacing are determined (see Ref.~\cite{klostermann_construction_2022} for further information).

\emph{Image rotation and phase extraction.--}
The reconstruction scheme based on the autoencoder architecture introduced in the main text (Fig.~\ref{fig:autoencoder_architecture}a) requires that the lattice axes are roughly aligned with the image edges due to the square occupation matrix in the bottleneck layer. Since our lattice axes enclose an angle on the order of $\SI{30}{\degree}$ with the edges of the image, we rotate the full images prior to reconstruction, using only linear interpolation to enable sub-pixel shifts. Additionally, our lattices deviate by $\lesssim\SI{2}{\degree}$ from a perfect square configuration, which however does not seem to degrade the reconstruction performance noticeably. The lattice vectors in the rotated frame are determined using the aforementioned procedure with rotated dilute images.

In order to fully determine the location of the lattice sites, we also require the lattice phase with respect to the image origin. This phase is obtained in each image individually by fitting the positions of isolated atoms that are always present in the vicinity of the MI.

\subsection{Data pre-processing}
\label{sec_sm:preprocessing}
\emph{Cropping of lattice segments.--}
To process fluorescence images of arbitrary size, we need to crop smaller segments containing exactly $16\times16$ lattice sites to be used as the input for the autoencoder network. Since the lattice spacing is not commensurate with the camera pixel spacing, we first up-sample the images by a factor of 20 without interpolation in order to avoid large discretization errors when cutting. This ensures that the lattice sites inside a crop are consistently located at the same positions relative to the origin. Finally, the cropped images are down-sampled to a size of $256\times256$ pixels using linear interpolation to fit the input layer shape of the autoencoder, where each lattice sites appears as $16\times16$ pixels.

\emph{Scaling of the training data.--} 
Scaling the input data before processing it using a neural network is essential for successful training. As a general rule, the magnitude of the input values should be on the order of one and symmetrically centered around zero. We re-scale the pixel values $z_{i}$ of the $16\times16$ site crops to the range $[-1,1]$ according to 
\begin{equation}
    z'_{i}=2\frac{z_{i}-z_{\mathrm{min}}}{z_{\mathrm{max}}-z_{\mathrm{min}}}-1,
\end{equation}
where $z_{\mathrm{min}}$ and $z_{\mathrm{max}}$ are the minimum and maximum pixel values in the entire training set, respectively. The scaling parameters $z_{\mathrm{min}}$ and $z_{\mathrm{max}}$ are determined from the training set and later used for re-scaling all new data subject to the reconstruction process.

Since non-linear activation functions are used throughout the architecture, the classification result becomes sensitive to the absolute scale of the input data. It is hence recommended to correct for varying counts by, for instance, measuring the counts of isolated atoms or monitoring the shape of the deconvolved count distributions and re-scaling the input data accordingly.

\subsection{Site counts in unprocessed fluorescence images}
\label{sec_sm:unproc_counts}
\begin{figure}[tbh!]
\begin{center}
  \includegraphics[width=0.99\columnwidth]{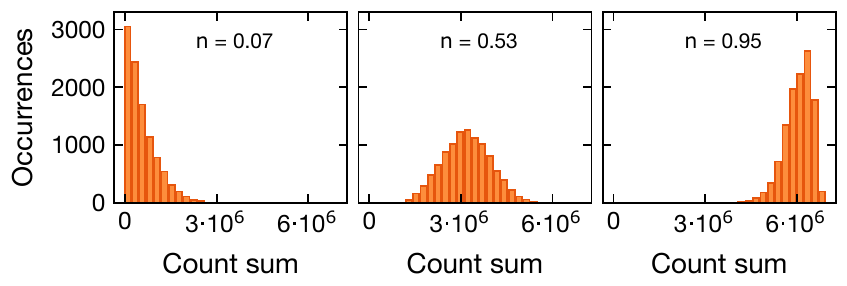}
  \caption{\textbf{Unprocessed count sum distributions.} Distribution of the count sums within each lattice site for images at three exemplary average filling values. There is no discernible separation, making a direct threshold discrimination impossible.}
  \label{fig:sm_exp_count_histo}
\end{center}
\end{figure}

\begin{figure*}[t]
\begin{center}
  \includegraphics[]{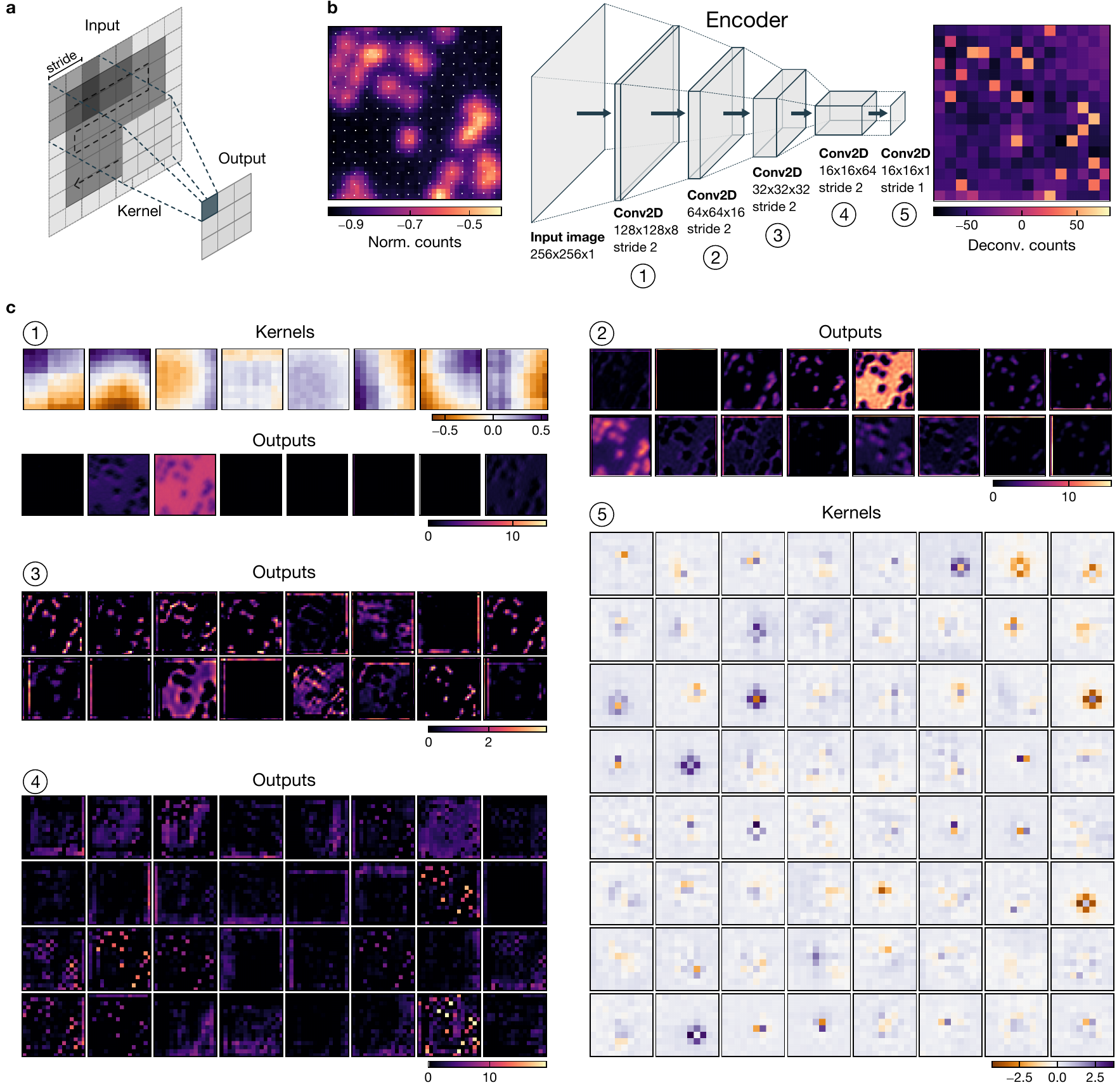}
  \caption{\textbf{Visualization of the convolutional operations in the encoder network.} \textbf{a} Schematic illustration of a convolution operation. A kernel with entries learned during training is moved across the input data, advancing at a fixed step width (stride). At each position, an inner product between the kernel and the input patch is computed. Shown here for a stride of two. \textbf{b} Visualization of the transformation applied by the encoder network for an experimental input image. \textbf{c} Selection of kernels and outputs of the different encoder layers. Note that we only show one half of the outputs for layers three and four due to the large number of kernels.}
  \label{fig:sm_encoder_vis}
\end{center}
\end{figure*}

In case of a smaller resolution ratio $\beta$, it is possible to directly discriminate empty and occupied lattice sites by means of a simple threshold reconstruction. In this method, the pixel sums of the lattice sites exhibit a bimodal distribution which allows to perform a threshold discrimination. As the ratio $\beta$ is increased, the spillover of a PSF onto the neighboring sites causes the bimodal distribution to wash out, rendering a threshold discrimination impossible. For $\beta=2.2$ in our experiment, the pixel sum distributions (examples in Fig.~\ref{fig:sm_exp_count_histo}) do not exhibit a discernible separation for any filling, highlighting the need for a powerful deconvolution algorithm.

\subsection{Visualization of the encoder network}
\label{sec_sm:encoder_vis}

\emph{Convolutional operations.--} The autoencoder architecture introduced in the main text (Fig.~\ref{fig:autoencoder_architecture}a) is constructed from several convolutional layers. The basic working principle is illustrated in Fig.~\ref{fig:sm_encoder_vis}a. An input image is traversed by a kernel, advancing by a given step width (stride). At each position, an inner product between the kernel and the input patch is computed, yielding one entry of the output matrix. A stride of one (together with appropriate padding) retains the shape of the input image, while a stride larger than one reduces the output shape. There are usually several kernels in each convolutional layer, resulting in a set of outputs.

\emph{Visualization of the learned kernels.--} Since the kernels encode the transformation learned by the network, a visualization of the kernels as well as the outputs of each layer can give further insight into the inner workings of the neural network. However, an interpretation of kernels in deeper layers is not straightforward as they are high-dimensional and operate on abstract features that have been pre-processed by previous layers. The development of methods to interpret neural networks is still an active field of research~\cite{zeiler_visualizing_2013}.

We investigate the transformation applied in the encoder by tracing the evolution of an experimental crop through the network~(Fig.~\ref{fig:sm_encoder_vis}b). We show a selection of kernels and outputs of the individual convolutional layers~(Fig.~\ref{fig:sm_encoder_vis}c). Since the first layer operates directly on the input image, a visualization of the kernels is straightforward. Here, we observe smooth structures in all eight learned kernels. The eight corresponding outputs show that no significant transformation happens at this stage yet, and the majority of kernels for this specific input image are suppressed. In the following three layers, we observe how a combination of both large scale and increasingly small-scale features is extracted. The last step before arriving at the deconvolved counts in the bottleneck layer (rightmost image in Fig.~\ref{fig:sm_encoder_vis}b) is a convolution with one 3D kernel at a depth of 64, corresponding to one 2D kernel for each output of the previous layer. Interestingly, these kernels exhibit structures where one central entry has a strong weight and the nearest neighbors are oppositely weighted. We interpret this shape to serve the purpose of reverting the spillover of counts from atoms on neighboring sites, which is a major factor complicating reconstruction at large resolution-to-spacing ratios.

\section{Evaluation on simulated data}
\label{supp_sec:simulations}

A central feature of the autoencoder architecture is the possibility of unsupervised training directly with experimental data, independent of a possibly imperfect simulation. Nevertheless, an evaluation of simulated fluorescence images can provide useful insights into the reconstruction algorithm due to the available knowledge about the actual occupation. Here, we simulate fluorescence images corresponding to arbitrary lattice occupations, reproducing our experimental conditions as closely as possible. Specifically, we use the extracted lattice vectors and the measured PSF (Fig.~\ref{fig:autoencoder_architecture}c), and we choose the photon counts per atom ($c_{\mathrm{atom}}, \, \sigma_{\mathrm{atom}}$) and background noise ($\sigma_{\mathrm{bg}}$) to match the measured SNR (Fig.~\ref{fig:autoencoder_architecture}d).

\emph{Simulation procedure.--} 
To create a simulated fluorescence image, we proceed as follows: (1) A grid of site coordinates is spanned according to the measured lattice vectors. (2) For each occupied lattice site, we draw a number of scattered photons from a normal distribution with mean and variance ($c_{\mathrm{atom}}, \, \sigma_{\mathrm{atom}}$). (3) The positions of the scattered photons are sampled according to the experimentally measured PSF, capturing asymmetries and long-range aberrations. (4) Photons are then binned according to the camera pixel grid and (5) the calibrated background noise level $\sigma_{\mathrm{bg}}$ is added per pixel.

Before training or reconstruction, simulated images are pre-processed as described in Section \ref{sec_sm:preprocessing} in the exact same way as experimental images.

\emph{Architecture and training.--} We use the simulated fluorescence images to train the network as described in the main text and using the composite loss function [Eq.~(\ref{eq:ae_loss_fn})]. Similar to the case of experimental data it is essential that the training dataset contains images at all fillings. For training, we use 70,000 simulated lattice sections.

\begin{figure}[tbh!]
\begin{center}
  \includegraphics[]{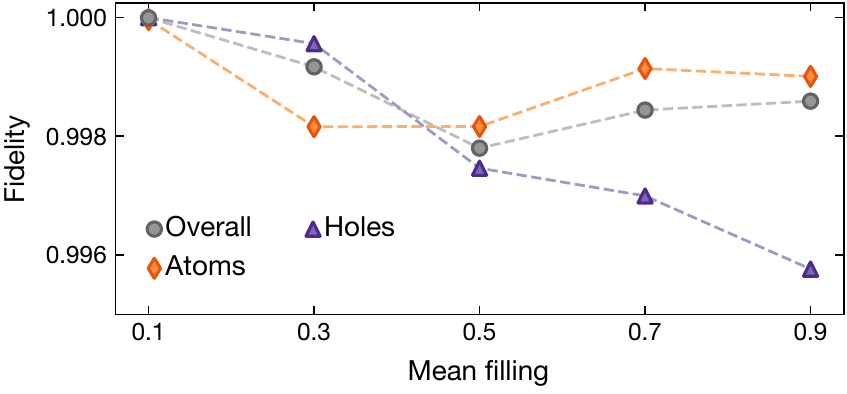}
  \caption{\textbf{Fidelity on simulated data as a function of the mean filling.} The fidelity has been evaluated on a test dataset not seen during training with about 1,000 crops per filling bin. The overall fidelity is the fraction of correctly classified sites, while the individual (hole) atom fidelities are the fractions of correctly classified (holes) atoms. Dashed lines are a guide to the eye.}
  \label{fig:sm_ae_sim_fidelity_vs_filling}
\end{center}
\end{figure}

\emph{Fidelity.--} 
The fluorescence simulation enables a straightforward quantitative evaluation of the reconstruction fidelity by comparing the network prediction to the true occupation labels. As in the experimental case, this is done using a previously unseen test set. We define the overall reconstruction fidelity as the ratio of correctly classified sites to the total number of sites $\mathcal{F} = {N_{\mathrm{correct}}}/{N_{\mathrm{tot}}}$. Additionally, it is interesting to separately evaluate the atom and hole fidelity to gain information on systematic errors in the reconstruction process, where we define ${\mathcal{F}_{\mathrm{holes}} = ({\mathrm{\# \, correctly \, predicted \, holes}})/({\mathrm{\# \, true \,  holes}})}$ and accordingly for atoms (occupied sites).

Fig.~\ref{fig:sm_ae_sim_fidelity_vs_filling} shows the reconstruction fidelities as a function of the average filling per image. While the overall fidelity has a minimum around half filling, the hole fidelity monotonically decreases toward higher fillings. This indicates that it is most difficult, even for simulated data, to identify individual holes in high-density regions. All fidelities stay well above \SI{99}{\percent} for all fillings, therefore exceeding the experimentally determined value in the main text (Fig.~\ref{fig:multi_img_fidelity}). The striking difference to the experimental results most likely comes from additional effects that have not been considered in the simulation. Examples are hopping and loss, density-dependent effects due to superradiance as well as other imaging imperfections such as spatially inhomogeneous imaging performance. 

\begin{figure}[tbh!]
\begin{center}
  \includegraphics[]{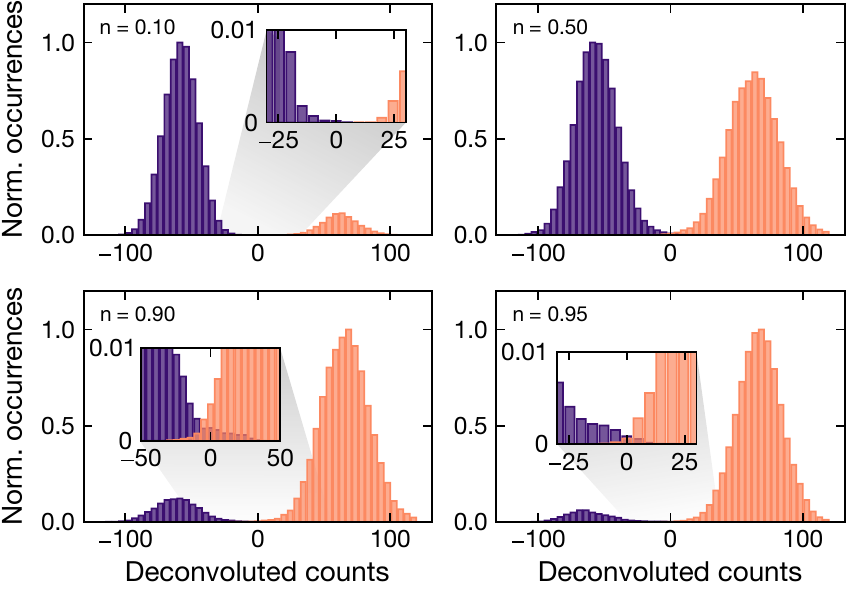}
  \caption{\textbf{Deconvolved site count distributions for different mean fillings.} Based on the true occupation labels, purple and orange mark empty and occupied lattice sites, respectively. Across all fillings, one finds a bimodal distribution with negligible overlap.}
  \label{fig:sm_ae_sim_histograms}
\end{center}
\end{figure}

\emph{Deconvolved count statistics.--} In addition to the quantitative fidelity evaluation, the simulation also provides deeper insights into the interpretation of the deconvolved count distributions. Similar to the evaluation of experimental data in the main text, we extract the values of the bottleneck layer before binarization via the activation function and show the distributions for different average fillings. In Fig.~\ref{fig:sm_ae_sim_histograms} one finds well-separated bimodal distributions for all fillings. The two peaks can be attributed to occupied and unoccupied sites based on the known occupation labels. In the vicinity of the intersection point where the two distributions overlap, one finds that the tails extending into the respective other class decay rapidly. This suggests that the overlap area can be used as an approximation to estimate the reconstruction fidelity also in the case where the underlying distribution is not exactly known. Compared to the experimental case (Fig.~\ref{fig:amp_histograms}), the qualitative shape of the distributions obtained here is much closer to a normal distribution. The strong skewing in Fig.~\ref{fig:amp_histograms} of the main text is thus likely to be related to hopping and loss as well as density-dependent effects, which are not included in our simulation.

\section{Fidelity Estimation From Double Imaging}
\label{supp_sec:double_imaging}

One of the two ways introduced in the main text for evaluating performance of the neural network reconstruction is imaging the same atom distribution twice, similar to the procedure for determining the loss and hopping rate, as described in Section~\ref{sec:suppfluo}. In the following, we will elaborate on how an experimental reconstruction fidelity can be derived from comparing the reconstruction results of the two images. In doing so, we assume that any errors in the reconstruction are caused by a finite SNR in presence of (random) noise and disregard systematic errors which entail a reconstruction bias toward either occupied or unoccupied sites.

\begin{figure}[tbh!]
\begin{center}
  \includegraphics[width=0.99\columnwidth]{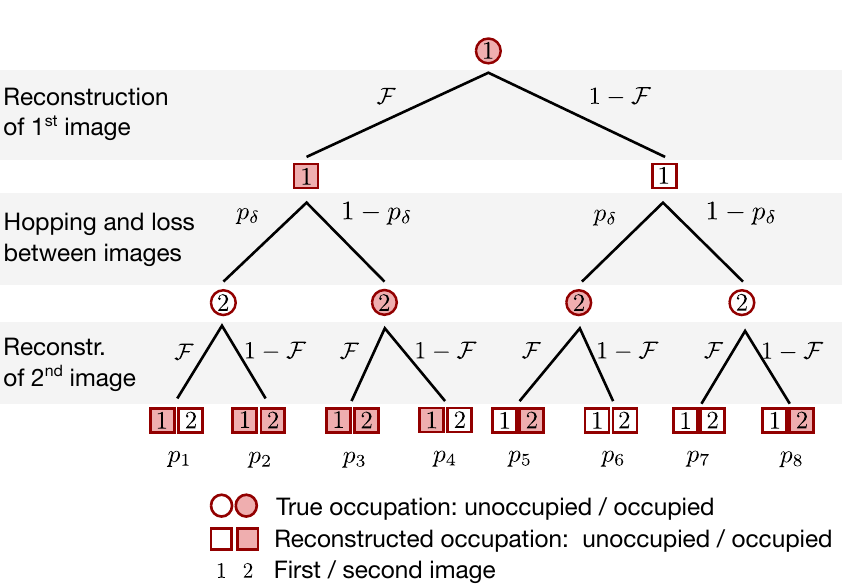}
  \caption{\textbf{Double imaging probability tree.} Probability tree modeling the reconstruction and hopping/loss processes as a three-step Bernoulli trial that depends on the reconstruction fidelity $\mathcal{F}$ and the hopping/loss probability $p_{\delta}$. Quadratic and round symbols (red/white color) show the true and reconstructed occupation (occupied/unoccupied), respectively, of a site in the first or second image (1,2). The probabilities of the eight outcomes of the probability tree are labeled $p_1, \dots, p_8$.}
  \label{fig:supptree}
\end{center}
\end{figure}

\emph{Bernoulli trial.--} Assuming that the hopping and loss probabilities are not too large, we can envision the process of taking two fluorescence images of the same cloud as a three-step process which can be thought of as a Bernoulli trial: (I) taking and reconstructing the first image, (II) hopping and loss causing actual differences between the two images and (III) taking and reconstructing the second image. Let us consider an arbitrary single site. The reconstruction fidelity $\mathcal{F}$ is the probability that a site in the first image is reconstructed correctly, while there is a probability of $(1-\mathcal{F})$ that the reconstruction yields the wrong occupation. Due to thermal hopping and loss events, the true occupation of the same site in the second image has a chance of $p_{\delta}$ to be different from that in the first image. The probability $p_{\delta}$ is proportional to both the average density in the image $n$ and the hopping and loss probabilities in the exposure time window ($p_{\mathrm{ loss}}$ and $p_{\mathrm{hop}}$, respectively):

\begin{equation}
    p_{\delta}(n) = n\,( p_{\mathrm{loss}} + 2\,p_{\mathrm{hop}})
\end{equation}

The factor of two in front of the hopping term accounts for the fact that a hopping event always alters the occupation of two sites, leading to a two-fold higher difference between the first and the second image as compared to a loss event. Finally, the probability that the site occupation (altered by loss and hopping) is reconstructed correctly in the second image is again given by the reconstruction fidelity $\mathcal{F}$.

The probabilities governing this three-step process can be summarized in a tree diagram as shown in Fig.~\ref{fig:supptree}. At the bottom of this diagram all eight outcomes with probabilities $p_1, \dots, p_8$ are shown. As an example, ${p_1=\mathcal{F}\;p_{\delta}\;\mathcal{F}}$ is the probability for the case where both images are reconstructed correctly but the two occupations are different due to a hopping or loss event. The probability $p_2=\mathcal{F}\;p_{\delta}\;(1-\mathcal{F})$ corresponds to the case where the two reconstructed occupations match even though there was a hopping or loss event, as the second reconstruction result is incorrect.

\emph{Fidelity estimation.--} Among the eight outcomes shown at the bottom of the probability tree only four (those numbered 1, 4, 5 and 8) cause a measured occupation difference between the first and the second image. Thus, the probability $\delta$ that the measured occupation of a site in the second image is different compared to the first image is:

\begin{align}
\delta &= p_1 + p_4 + p_5 + p_8 \nonumber \\ &= \mathcal{F}^2\,p_{\delta}(n) + \mathcal{F}\,(1-\mathcal{F})\,(1-p_{\delta}(n)) \nonumber \\ & + (1-\mathcal{F})\,\mathcal{F} \, p_{\delta}(n) + (1-\mathcal{F})^2\, (1-p_{\delta}(n)).
\label{eq:delta}
\end{align}
Solving this equation for $\mathcal{F}$ yields:
\begin{equation}
\mathcal{F} = \frac{1}{2} \left( 1 + \sqrt{\frac{1-2 \delta}{1-2\,p_{\delta}(n)}} \right).
\label{eq:rho}
\end{equation}
For vanishing hopping and loss rates ($p_{\delta} = 0$), this expression simplifies to:

\begin{equation}
\mathcal{F} = \frac{1}{2} (1 + \sqrt{1-2 \delta} ).
\label{eq:rho_simple}
\end{equation}

If no difference between the first and the second image is measured ($\delta = 0$), we obtain a perfect reconstruction fidelity of $\mathcal{F} = 1$. In contrast, when measuring the maximum possible difference $\delta = 0.5$ (which is equivalent to comparing two random images), Eq.~(S5) yields $\mathcal{F} = 0.5$, the lowest possible value for the reconstruction fidelity.

\putbib[references_supplement]
\end{bibunit}